# SYNTHETIC KNOWING: THE POLITICS OF INTERNET OF THINGS

**Authors**

Eric Monteiro, Department of Computer Science, Norwegian University of Science and Technology (NTNU), NO-7491 Trondheim, Norway, eric.monteiro@ntnu.no

Elena Parmiggiani, Department of Computer Science, Norwegian University of Science and Technology (NTNU), NO-7491 Trondheim, Norway, parmiggi@ntnu.no

**Author biographies**

Eric Monteiro is professor of Information Systems at the Norwegian University of Science and Technology (NTNU) and adjunct professor at the University of Oslo. Drawing on an interdisciplinary perspective influenced by Science and Technology Studies, he is interested in the digital transformation in organizations of work and knowing, notably in healthcare and energy/oil. He has worked extensively with a theoretical lens of large-scale change projects informed by (information) infrastructure/digital platforms. His publications have appeared in MISQ, ISR, JAIS, Information and Organization, EJIS, The Information Society, Science, Technology & Human Values, and CSCW Journal.

Elena Parmiggiani is postdoctoral researcher in Information Systems at the Norwegian University of Science and Technology (NTNU) and holds a PhD in Information Technology from NTNU (2015). She is interested in studying the sociotechnical challenges of implementing, integrating, and maintaining digital platforms and information infrastructures, and in the methodological stakes of studying distributed and long-term arrangements. Her work is based on an interdisciplinary lens influenced by Science and Technology Studies. Empirically, she has focused on environmental monitoring and the oil industry. She has published primarily within Information Systems and CSCW.

**Abstract**. All knowing is material. The challenge for Information Systems (IS) research is to specify *how* knowing is material by drawing on theoretical characterizations of the digital. Synthetic knowing is knowing informed by theorizing digital materiality. We focus on two defining qualities: *liquefaction* (unhinging digital representations from physical objects, qualities, or processes) and *open-endedness* (extendable and generative). The Internet of Things (IoT) is crucial because sensors *are* vehicles of liquefaction. Their expanding scope for real-time 'seeing', 'hearing', 'tasting', 'smelling', and 'touching' increasingly mimics phenomenologically perceived reality. Empirically, we present a longitudinal case study of IoT-rendered marine environmental monitoring by an oil and gas company operating in the politically contested Arctic. We characterize synthetic knowing into four concepts, the former three tied to liquefaction and the latter to open-endedness: (i) the objects of knowing are algorithmic phenomena; (ii) the sensors increasingly conjure up phenomenological reality; (iii) knowing is scoped (configurable); and (iv) open knowing/data is politically charged.

# 1 INTRODUCTION

Under the banner of knowledge management, Information Systems (IS) research over the past few decades has observed a surge of interest in how knowledge can be understood and conceptualized (Alavi and Leidner 2001). Driven by deep-seated trends in the social sciences, practice-based perspectives on knowing underscore the embodied, embedded, and emergent aspects of knowing (Blackler 1995; Feldman and Orlikowski 2011). As Orlikowski (2006, p. 460) notes, practice-based perspectives are perfectly well aware of the fact that all knowing involves materiality, yet on "the level of conceptualization and theorizing, we tend to disregard this". There is accordingly a need, Orlikowski (ibid.) argues, to supplement the practice-based perspectives on knowing with an explicit conceptualization of materiality. She makes this a general observation about practice-based perspectives and cites examples of modest (e.g., the possibility to mute the microphone during a video-conference) and moderate (e.g., creating an electronic archive that is available to all users regardless of geography) entanglement of materiality in the knowing. When, as in our case, the role of materiality in knowing is *constitutive* rather than modest or moderate, this need for a conceptual supplement to practice-based knowing is significantly compounded.

The entanglement of material knowing is highlighted by extreme cases of digitalization. This is because a defining capacity of digitalization is the 'liquefaction' or unhinging of the digital representation "from its related physical form or device" (Lusch and Nambisan 2015, p. 160; cf. also Barrett et al. 2015). This entails that the objects of knowing are increasingly 'self-referential' digital representations rather than physical objects or processes (Kallinikos 2007). In this context, the Internet of Things (IoT) takes on a particularly important meaning. First, sensors *are* vehicles for liquefaction, as they generate digital representation inferred from physical objects, properties, and processes (Oreskes et al. 1994). As the number and types of sensors expand, the types of physical phenomena amendable to liquefaction expand. Second, knowing presupposes human sensory capacity as Zuboff (1988, p. 62) noted "I see, I touch, I smell, I hear; therefore, I know". Human sensory capacity, however, is increasingly being approximated by sensors hence not in the same manner as when Zuboff wrote an exclusively human capacity.

Thus, Orlikowski's (2006) general call for theoretical attention to material knowing is significantly magnified with extreme cases of digitalization: what we know (digital representations) and how we know it (algorithmic manipulation) are constituted by digitalization.

Digitalization qua liquefaction challenges how, if at all, digital representations become practically 'real'. A promising theoretical approach is Knorr Cetina's (2009) notion of a synthetic situation. It builds on the firmly established situated nature of knowing underscored by practice-based perspectives. Knorr Cetina defines synthetic situations as augmented environments in which specific digital representations (in her case, numbers, graphs, and charts) constitute financial traders' situations (ibid., p. 69). Synthetic situations are as 'real' as physical, local "particular, concrete circumstances" (Suchman 2006, p. 26). What Knorr Cetina offers is that the situated knowing of practice-based perspectives is theoretically extended into synthetically situated knowing (in short, synthetic knowing) when the material entanglement of knowing is constitutive rather than modest or moderate. Knorr Cetina (2009) provides a useful point of departure for detailing material knowing informed by theorizing digital materiality, i.e., synthetic knowing.

We draw on a longitudinal case study of the ongoing efforts of an oil and gas operator to establish a capacity for real-time, IoT based marine environmental monitoring in the Arctic region in northern Europe. Digitalization is constitutive for material knowing as the marine environment (the object of knowing) is increasingly liquefied and hence transformed from its physical and biological basis into sensor-based, digital representations. Hence, objects of knowing are shifted from, e.g., physical schools of fish to acoustically sensor-based, computationally manipulated object of 'biomass'. The methods/tools of knowing are increasingly model- and data-driven with algorithmic manipulations and digital visualization. This transforms the sampling-based detection of pollution of the marine environment from, e.g., the dispersion of rock cuttings generated during oil well drilling into a digitally visualized model- and sensor-driven prediction. The Arctic is inherently politicized. The Arctic is of significant commercial interest to an oil and gas industry under increasing economic pressure. Simultaneously, they are extremely rich fishing grounds and sites of natural beauty. Synthetic knowing represents a new mode of knowing likely to grow in relevance yet to date under-researched and central to the politicized knowing of the Arctic. Responding to Orlikowski's (2006) call, our research question is "How is synthetic knowing theoretically characterized in IoT-saturated contexts?"

The remainder of this paper is organized as follows. Section 2 provides our theoretical framework. Synthetic knowing is informed by theorizing digital materiality (Nambisan et al. 2017; Orlikowski and Iacono 2001). We discuss two defining qualities: *liquefaction*, outlined above, and the generative (Zittrain 2006), extendable (Kallinikos 2007), programmable (Yoo et al. 2010) or *open-endedness* of digital materiality. These two qualities of digital materiality are prerequisites synthetic knowing. The open-endedness quality in our case focuses on 'open' knowledge (data, 'facts') and is subject to ideological and rhetorical strategies (Barrett et al. 2013). Section 3 provides the case context, section 4 the methods, and section 5 presents the case narrative of our longitudinal (April 2012–December 2017) case study at the oil and gas company (dubbed NorthOil for anonymity). The discussion in section 6 elaborates the notion of synthetic knowing, our principal theoretical contribution. We characterize four aspects of synthetic knowing; the former three tied to liquefaction, whereas the last relates to the open-endedness of digital materiality: (i) the objects of synthetic knowing become *algorithmic phenomena*; (ii) real-time sensor streams increasingly mimic the conjuring up of *phenomenological reality*; (iii) synthetic knowing is in the formulation of Knorr Cetina *scoped*, i.e., configurable; and (iv) 'open' data/knowing, played out through ideological and rhetorical strategies, is *political*. Section 7 offers concluding remarks including reflections on the boundary conditions for synthetic knowing.

## 2 PERSPECTIVES ON KNOWING

### 2.1 Material Knowing: On Being Specific About the Digital

Anything but reified, knowing underpins action/practice (Alavi and Leidner 2001; Styhre 2003). Knowing, Orlikowski (2006, p. 460, emphasis in original) notes, is "*emergent* (arising from everyday activities and thus always 'in the making'), *embodied* (as evident in such notions as tacit knowing and experimental learning), and *embedded* (grounded in the situated socio-historic contexts of our lives and work)". Hence, knowing is "a situated knowing constituted by a person acting in a particular setting and engaging aspects of the self, the body, and the physical and social worlds" (Orlikowski 2002, p. 252).

However, as influential insights in the social sciences have made clear during the last couple of decades, *all* knowing practices are material (Latour 1999). This insight forms the backbone of the broad research program on sociomateriality (Cecez-Kecmanovic et al. 2014; Orlikowski and Scott 2008).

There is thus a broad consensus *that* knowing is material, but significant divergence regarding *how*, be it 'entangled' (Orlikowski 2006), 'imbricated' (Leonardi 2013), 'inscribed' (Hanseth and Monteiro 1997) or 'performative' (Mackenzie 2006). Rather than pursue the general agenda of sociomateriality that "matter matters" (Barad 2003), we respond to calls to theorize the specifics of digital materiality rather than to apply generic theory (Nambisan et al. 2017; Orlikowski and Iacono 2001). So how, then, to conceptualize digital materiality? Proposals abound. As vehicles for our subsequent analysis, we identify two salient qualities of digital materiality that tap into deep-seated themes in earlier conceptualizations.

First, there is the defining capacity of digital representations for liquefaction or unhinging from the physical objects, processes or properties (Barrett et al. 2015; Lusch and Nambisan 2015). This resonates, despite differences in vocabulary and emphasis, with important insights by scholars. Liquefaction corresponds to the 'virtual' quality (or rather: potential) of digital representations to 'stand in for' reality (Bailey et al. 2012). Similarly, liquefaction is what allows the 'layered' quality of digital technologies (Yoo et al. 2010), the same quality that enables the 'recombinations' at the core of digital innovation (Henfridsson et al. 2018) or service-dominant logic (Lusch and Nambisian 2015). The capacity for liquefaction is what underpins the 'fluidity' of the virtual/physical distinction in Knorr Cetina's (2009) synthetic situations hence their potential for 'immersion'.

Second, we focus on the characteristic of the digital as inherently open-ended. This quality too taps into recurring themes in efforts to conceptualize the digital. Zuboff's (1988) early and influential work analyzed what specifically was different with digital technologies compared to other technologies. In her analysis, the distinctive quality of digital technologies was their capacity to also 'informate' in addition to what all technologies (digital ones included) do, 'automate'. The capacity to informate is the open-ended capacity for repurposing, aggregation, and further manipulation stemming from the fact that digital representations are not consumed but "renders events, objects, and processes…visible, knowable, and shareable in a new way" (ibid., p. 9). More recently, this has been reiterated in alternative but related formulation of characterizations of the digital as generative (Zittrain 2006), incomplete (Garud et al. 2008), programmable (Yoo et al. 2010) or editable (Kallinikos 2007). Committed to the relevance to material knowing, we discuss how this open-ended character of the digital feeds and frames rhetorical and ideological strategies for 'open' knowing (see, e.g., Gillespie 2010; Ruppert 2015).

**2.2 Synthetic Knowing**

The liquefaction capacity of digital representations immediately threatens the dichotomous separation between the real vs. virtual/digital. This dichotomy, Boellstorff (2016, p. 388) notes, fails to address the principal challenge in accounting for material knowing in a digital setting viz. accounting for "precisely how the digital can be real" (cf. also Bailey et al. 2012).

Knorr Cetina's (2009) notion of a synthetic situation is a helpful means to dismantle the dichotomy of real vs. virtual/digital. The notion of synthetic situation grants the same status to the real and the digital with regard to their role in practices of knowing. In her case, Knorr Cetina (ibid.) portrays the transformation of financial trading from physically co-located settings to distributed and electronically mediated settings. As an empirical phenomenon, "a 'situation' invariably includes, and may in fact be entirely constituted by, on-screen projections" (ibid., 65). The synthetic situation "stitches together an analytically constituted world made up of 'everything' potentially relevant to the interaction" (ibid., 66). Empirically including digital representations is fundamentally different from the symbolic interactionalist's definition of a situation, which, "despite nods…was, at its core…a physical setting or place" (ibid., 63).

What synthetic situations teaches us, then, is that liquefied, digital representations may be as real as the physical in material knowing. Our principal interest, motivated by the characteristics of the empirical case, is with material knowing in synthetic situations where the virtual/digital representations dominate rather than merely supplement. We dub this synthetic knowing. Here IoT based renderings of reality are particularly important. Sensors—the quintessential generators of liquefaction—expand in reach and scope the type of phenomenon made synthetically knowable. Different from simulations (Dodgson et al. 2007), the *real-time* renderings of reality that the IoT allows, also evident in the ticker streams of traders in Knorr Cetina's case (2009), feed the ongoing perception hence are particularly compelling in making the virtual 'real' (Vertesi 2012). Expanding in scope and reach, sensors increasingly mimic embodied perception ('seeing', 'hearing', 'tactile sensation', 'smelling', and 'movement/balance', cf. Singh et al. 2014). Understood phenomenologically, sensors allow for sensing the lifeworld in already interesting but rapidly expanding richness that increasingly approximate human sensory capacity.

**2.3 Towards Synthetic Knowing: political perspectives**

Knowing, as scholars remind us, is inherently caught up with organizational politics. Perceptions, interests, and agendas among the actors, groups, and stakeholders vary. Knowledge/knowing is accordingly anything but neutral. As Alvesson (2001, p. 866) notes, the politics of knowing may result in knowledge of certain types or held by certain actors being privileged, e.g., when managers "downplay the role of technical expertise" in a tech company. Still, elaborate accounts of the forms and implications of political aspects of knowing are "less addressed from current research in knowledge phenomena" (Asimakou 2009, p. 84).

A productive approach to the politics of knowing is to analyze ideological and rhetorical strategies implicated in technology development, use and diffusion drawing on the work of Swanson and Ramiller (1997). Particularly relevant to our focus on material knowing of digital phenomena is the work on ideology and rhetoric tied to 'openness' (of data, knowledge/facts, code) because it taps directly into the second quality of digital materiality identified above (i.e., open-endedness).

Conceptualized in line with Giddens' broader structuration theory (Jones 2014), Barrett et al. (2013) elaborate how ideology acts as (Giddens') deep structures whereas rhetorical framing (textual or other representations) are the visible expressions (akin to Giddens' actions). We follow Barrett et

al.'s (2013, p. 217) invitation to draw on ideology/rhetorical framing of openness to "explore further these economic and political dimensions more explicitly in relation to the diffusion of technological innovations, and the discursive argumentations and justifications accompanying them". Drawing on Barrett et al. (2013), Shaikh and Vaast (2016) demonstrate how the ideologically espoused openness in an open-source project was undermined in practice. Rhetorically, the authors hold, openness was heralded, whereas, pragmatically, open-source developers navigated their way through semi-closed arenas fluidly. Similarly, and more relevant to our case, the means of opening up data/knowing through 'platformization' is subject to rhetorical and ultimately political maneuvering (Ruppert 2015).

In summary, synthetic knowing responds to the general call that knowing is material (Orlikowski 2006, p. 461) by zooming in on the specific materiality of the digital. Unhinged or liquefied from their physical referents, digital representations *might* become as real as the physical, and the touted open-endedness *might* become real. However, what, if any, weight digital renderings of reality carry remains an empirically under-researched question (Bailey et al. 2012). Our case of an IoT based, real-time rendered marine environment in the oil-rich yet extremely vulnerable Arctic region provides a vivid testbed for synthetic knowing.

## 3 CASE CONTEXT

The Arctic is a pristine marine environment. Rich in flora and fauna, it is home to abundant fishing grounds. An estimated 25% of unexplored oil and gas worldwide is also in the Arctic (Bird et al. 2008). However, knowledge of the Arctic is limited. The Arctic is accordingly subject to heated and ongoing political debates. Controversies include diverging assessments of the operational safety (e.g., harsh weather and ice) and the financial and environmental risks posed by oil and gas activities.

There is a wide spectrum of stakeholders, including governmental and public agencies, industrial lobby groups, corporate interests, labor unions, and environmental activists. These stakeholders actively contribute to generating, supporting, and legitimizing knowledge of the marine environment in the Arctic. Shifting from mere consumers, oil companies are increasingly engaged in the production of data about the environment. NorthOil (a pseudonym) is one of the major oil companies in Norway and operates on all continents worldwide. Since the early 2000s, NorthOil has strengthened its technological and scientific capacity for IoT based marine environmental monitoring, with the objective of shifting from a corrective, ex post facto approach to a preventive, real-time approach for assessing environmental risk. Several, including environmental activists and fishermen, challenge the knowledge produced by oil companies (Lamers et al. 2016).

There is no lack of controversies. Even what constitutes oil and gas 'operations' is debated. For instance, seismic surveys (shooting bursts of acoustic sound waves at the geological formations below the seabed and reading off the echoes) might be allowed, but, as environmentalists and fishermen have noted, they still disturb whales and other sea life with unknown long-term effects[1]. Moreover, the same seismic 'data' can be interpreted differently: for oil operators, such data measure the 'viability' of their activities, whereas for fishermen, they indicate the 'vulnerability' of fish stock (Blanchard et al. 2014).

Political controversies regarding oil are hardly new in Norway. Throughout its 50 years of history with offshore oil, controversies abound. A principal means to break out of political dead-lock has been to commit to an ostensibly knowledge-based approach to political decision-making (Norwegian Ministry of the Environment 2011). Keeping certain areas off limits for oil activities has been a key mechanism for negotiating the inherently contradictory interests related to oil activities and environmental and fishing needs. For instance, until 1979, Parliament imposed a ban on oil activities north of 62° latitude due to the perceived risks to safety and the environment.

---

[1] Vegstein, L. U. 2014. "Kaller Lofoten-Vern En Bløff [Calls Lofoten Conservation a Bluff]," Klassekampen (July 12).

We focus on the ongoing political controversies surrounding bans on oil and gas activities in two areas of the Arctic: the Lofoten-Vesterålen-Senja area (dubbed Venus by us) and a particular area of the Norwegian part of the Barents Sea. We analyze how synthetic knowing ('facts') about the two areas are produced and challenged as part of negotiating the bans.

The former area, Venus, is currently off limits but is commercially interesting for oil operators and is among the world's richest fishing grounds for cod and herring. Venus has been a principal source of food and revenues for Norwegians for centuries. The IoT-based synthetic knowing in Venus evolves around estimating the amount of fish ('biomass') against a baseline of natural variation.

The ban on oil and gas activities in the latter area was lifted in 2013 despite outcries from environmental NGOs: "Our comment is that this is not knowledge management. Here we are in a situation where science adapted to politics, whereas it should have been vice versa"[2]. Located further north than Venus, the IoT-based synthetic knowing in this area evolves around the ice edge (i.e., the border of permanent ice) because of the operational and environmental hazards. The definition of the ice edge is highly controversial and debated. If the ice edge is defined further north than its present stipulation, as noted by one environmental NGO[3], "Norway would drill [for oil] farther north than anyone else…closer to the ice edge, further offshore, in extremely productive biologically and, hence, vulnerable areas…[with] increased likelihood and consequences of accidents".

---

[2] See Lorentzen, M. 2015. "Høring Om Oljevirksomheten i Barentshavet: Oljebransjen Advarer Om at Russland Kan Komme Oss i Forkjøpet [Hearing on Oil Business in the Barents Sea: The Oil Sector Warns That Russia Could Get There before Us]," E 24. (http://e24.no/energi/hoering-om-oljevirksomheten-i-barentshavet-oljebransjen-advarer-om-at-russland-kan-komme-oss-i-forkjoepet/23450809). Accessed May 18, 2016.

[3] News bulletin from Bellona, a high-profile environmental NGO in Norway, published a week after the Paris climate agreement. http://bellona.no/nyheter/olje-og-gass/2016-05-regjeringen-ofrer-iskanten, accessed May 18, 2016.

The principal data for this paper are fieldwork (see details below) following two of NorthOil's efforts to establish IoT-based synthetic knowing of the Arctic. The Venus Ocean Observatory (Venus project for short; a pseudonym) involves several initially independent, small-scale efforts since 2005 to develop and install ocean sensor networks in the Venus area. The EnviroTime project (a pseudonym) was a corporate initiative for environmental monitoring run in 2011-2014, headed by NorthOil in collaboration with a consortium of industrial partners.

## 4 RESEARCH METHODS

### 4.1 Access to the Case

This paper presents a longitudinal (April 2012–December 2017) case study of IoT based marine environmental monitoring in the context of offshore oil and gas operations. Oil and gas companies are secretive. Access to empirical data is difficult, as indicated by the limited number of case studies published in IS journals (for an exception, see Hepsø et al. 2009 and Østerlie et al. 2012). Our access to NorthOil and its partner companies was the result of prolonged engagement. The first author has a series of research collaborations with NorthOil spanning two decades. Access to the EnviroTime and Venus projects required negotiation. A member of our research network holds a position in NorthOil's research and development department. This member introduced us to the person responsible for the EnviroTime project. In April 2012, the second author was granted access (signaled by an employee badge) to NorthOil premises as a guest. After a few months, we were made aware of the Venus project. As a relevant alternative to the EnviroTime project to marine environmental monitoring, we followed also this initiative. By participating in EnviroTime and Venus project meetings, the two authors met and were subsequently granted access to four NorthOil project partner companies: two marine sensing technology vendors, one marine research institute, and one international organization for risk assessment methodologies and quality certification. Our case selection focusing on the EnviroTime and Venus project had marked elements of opportunistic maneuvering. Against a backdrop of our extended history of research collaboration with NorthOil, however, we theoretically were attracted to the projects as interesting, ambitious IoT-based initiatives.

**4.2 Data Collection**

We draw on three data sources: participant observation, semi-structured interviews, and documents. See Table 1 for a summary.

First, consistent with an ethnographically oriented approach, the primary data source was participant observations from April 2012 to December 2014. Participant observation was conducted almost entirely by the second author, albeit at times with the first author. The second author spent an average of three full working days per week in the field for the first two years and approximately two days per week during the third year. The second author had a shared desk in an open office space with key participants in the EnviroTime project. She observed daily activities and discussions, also over coffee and lunch breaks. This allowed the authors to take part in meetings and workshops that were held both internally at NorthOil and, later, with the partner companies.

Over time, the prolonged presence of the second author at NorthOil led to an increase in trust with the EnviroTime informants. She was given small tasks, such as commenting on memos or minutes from meetings. Similarly, she helped testing the EnviroTime software for real-time visualization of environmental risk and helped exploring new functionality added to the Venus web portal. In this manner, she was gradually drawn into internal NorthOil discussions, including candid ones, about EnviroTime. Throughout this process, we were sensitive to the different, sometimes conflicting concerns and interpretations expressed by different actors. Nevertheless, the good relationships with key informants were vital, as they gave us access to data that would be otherwise inaccessible.

Second, semi-structured interviews overlapped in time with participant observations. Most were conducted by the second author, but the first author joined for some. The interviews traced out experiences with and perceptions of the EnviroTime and Venus projects, including their understanding of the political debates. The interviews were facilitated by participant observations in the sense that the informal exchanges with the EnviroTime project participants while traveling by plane, train, or taxi allowed for a snowballing strategy for identifying new informants. After first interviewing members of the partner companies at NorthOil premises, we were gradually allowed to conduct interviews at the partner companies without being accompanied by NorthOil representatives.

Third, we collected two types of documents, internal and public. The second author collected the former, whereas the latter were collected by both authors. Access to internal documents (project memos, reports, presentations, and email exchanges) relied heavily on the gradual trust established with NorthOil informants. Locating relevant information in NorthOil's intranet, for instance, is like finding the proverbial needle in the haystack and prohibitive without help. Public documents consisted of regulations, policy and white papers by national authorities. We also collected news articles, Norwegian in addition to UK and US, relating to oil and gas activities in the Arctic up till December 2017. Public documents were particularly useful in pursuing politically contested topics, such as determining the ice edge. They were resources that we drew on in the interviews and participant observations, too. Equipped with news clips, we sought comments and reflections from our informants.

Table 1. Overview of the three sources of data collection April 2012–December 2017

| **Participant observations** Approximately 1,200 pages of field notes | - shared NorthOil office space April 2012–December 2014<br>- 49 meetings, workshops, seminars, briefing sessions<br>- 36 project meetings (regular briefing sessions)<br>- 14 other events (conferences, seminars, workshops) related to but independent from project events |
|---|---|

| **Semi-structured interviews** (38; transcribed; average duration of 1 h) | - 17 with NorthOil employees (engineers, environmental advisors) involved in the EnviroTime and Venus projects<br>- 4 with NorthOil data management experts<br>- 9 with environmental experts from a partner company<br>- 1 with the vice president of a partner company<br>- 2 with NorthOil computer engineers<br>- 3 with marine acoustics experts from a technology vendor partner<br>- 2 with marine acoustics experts at a marine research institute |
|---|---|
| **Document study** | NorthOil internal documents (2012-2015):<br>- Microsoft SharePoint project team sites in NorthOil<br>- EnviroTime and Venus reports, memos, presentations<br>- Selected email exchange internally in NorthOil and with project partners<br>- NorthOil intranet (reports, presentations)<br>- Venus and EnviroTime software platform with documentation<br>External documents (2012-2017):<br>- repositories of national regulation and policy papers<br>- 60 news clips from national and international media |

**4.3 Data Analysis**

Based on an interpretative research approach (Klein and Myers 1999; Walsham 2006), our initial orientation to data collection and analysis was explorative. The data analysis was iterative and overlapped with data collection, thus granting flexibility to respond to emergent themes (Eisenhardt 1989). It was conducted jointly by the two authors. Inevitably reductionistic, we deconstruct our data analysis into three parts for the purpose of reducing the "inherent creative leap" (Langley 1999, p. 691).

We started off broadly exploring the work practices and tools involved in marine environmental monitoring. Keenly aware of the contested nature of environmental aspects of oil and gas, we were conscious to use informants from the EnviroTime and Venus projects also to identify to supplementary data sources and stakeholders. Data were coded manually (using colors, annotations and Post-it notes). We developed descriptive codes capturing our informants' views and reflections about marine environmental monitoring. For instance, one such descriptive code, "Balancing what can be measured and what is needed", illustrates the cross-pressure of trying to capture 'everything' about the marine environment against more pragmatic concerns for what is technical and economical viable and relevant to NorthOil.

As we subsequently worked with the data and clustered descriptive codes, themes emerged. Adhering to the actor- or stakeholder-centric principle of interpretative research, we identified what was perceived as a problem, why, by whom and what, if any, solution was proposed. For instance, a NorthOil member during an EnviroTime project meeting asked, "What's the average amount of cod in this area each month?", thus voicing a concern of the fishermen. Together with efforts to develop a 'virtual sensor' that "summarizes the presence of biomass [in the water column]", this led to the concept of an IoT-based 'definition of biomass', a concept evolving around the concerns for the range of natural variation of fish and other selected marine species.

Finally, our inductive data analysis engaged with theoretical imports. Anything but clean slates (Suddaby 2006), our experience influenced the analysis. Our long-standing research interest in work practices, technology, and organizational change was what attracted us to the case of marine 'reality' and caused Knorr Cetina's notion of synthetic situation to resonate early and deeply with us. Our theoretical curiosity stirred, we simultaneously found her notion lacking in specificity. Our engagement with theoretical imports in the data analysis should be understood as our process to operationalize and further specify Knorr Cetina's notion.

Working inductively, we were struck by the manner in which the marine environment—fish, water quality, corals, and eggs, for example—was pushed into the empirical background while digital representations filled their role. Crucially, this triggered us to conceptualize qualities of digital materiality. The disembedding, empirically, not initially theoretically, of digital representations increasingly standing in for physical, marine environment almost literally corresponds to the capacity for liquefaction (Lusch and Nambisan 2015). As elaborated in our section above on theory, this capacity taps into IS research discourses, notably on the 'virtuality' of digitalization (Bailey et al. 2012, Kallinikos 2007) and the role of recombinations in digital innovation (Henfridsson et al. 2018, Yoo et al. 2010). In our resulting interpretative template characterizing synthetic knowing in Table 2, three of the four constructs are tied to inductively emerging aspects of liquefaction: the way in which digital representations, algorithmically manipulated, increasingly become the phenomenon, the efforts to mimic sensory experience with IoT, and the inherent limitations of what a sensor may recognize.

The fourth construct of our interpretative table is tied to the open-ended quality of digital representations, a quality underscored already by Zuboff (1988). Inductively, our interest with the quality of open-endedness was tied to the political connotations of an IoT-rendered Arctic. Open data/ knowing was instrumental in attempting to mobilize stakeholders, also antagonistic ones such as fishermen. In addition, the open data/knowing rhetorically draws heavily on assumptions, if not ideologies (Barrett et al. 2013), about neutral, value-free science underpinning the so-called knowledge-based policy making Norwegian authorities ostensively advocate in the Arctic. Our later Discussion section is organized to theoretically elaborate on our characterization of the four constructs of synthetic knowing.

Our presentation of the Findings draws on a temporal bracketing strategy employed during data collection and data analysis (Langley 1999). Key to temporal bracketing is how to delineate the empirical phases. They are constructs of data analysis rather than given. In our case, the phases result less from chronology than from tracing out how IoT-based marine environmental monitoring was made politically relevant: first, by making the case for IoT being a plausible alternative to traditional practices; then, by making it credible by supporting historic trends over time against a baseline; and lastly, by generalizing the role of IoT to Arctic areas further north.

Table 2. The interpretative template resulting from our data analysis

| Construct | Concept | Excerpts |
|---|---|---|
| The objects of synthetic knowing become algorithmic phenomena | Definition of the ice edge | "Norway redefines Arctic ice edge in potential boost for oil exploration." (Newspaper headline) <br> "If someone wants to develop a field [by the ice edge], you have to collect data over the long term to understand the biological consequences of [the drilling activity]." (Meeting, expert in acoustics, NorthOil) |
| | Definition of biomass | "What's the average amount of cod in this area each month?" (Interview, environmental advisor, partner company) <br> "We came up with the idea of adopting a biomass indicator based on a 'virtual sensor' that summarizes the presence of biomass [in the water column] at various intervals." (Environmental advisor, NorthOil) |
| Real-time streams conjure up phenomenological reality | Mimicking sensory experience | "[A cod suddenly comes in front of the subsea camera] And that's what happens, he gets really angry. So he says "Shshshshsh!" (…) Or maybe he gets annoyed." (NorthOil IT advisor looking at the live feeds from the Venus web portal during an interview) |

|  | Selecting IoT-rendered qualities of the marine environment | "Spawning fish produce sounds ('grunts') that are characterized by a low frequency (50-500 Hz) and include 'basic pulse' repeated in different patterns and speeds. Those 'grunts' can be recorded by hydrophones (passive acoustics) and analyzed to determine the presence of spawning fish in the area…Combining results from passive and active acoustic with other key sensor data will make it possible to determine the distribution and migration of fish in relation to temperature, salinity, and the abundance of zooplankton." (Internal project report draft) |
|---|---|---|
| Synthetic knowing is scoped | In/visibility of sensors | "[S]pecies like the mackerel which don't have a swim bladder will return a very [weak] signal. Perhaps that's why we have come up with species with a swim bladder in this project." (Interview, environmental advisor, partner company) |
|  | Reach vs granularity of sensors | "We can't… [measure] larvae and eggs in the upper water masses because this [sensor network] is on the bottom [at -250 m] and has a reach of about 50 meters." (Interview, environmental advisor, partner company) "In [Venus] there is only one [sensor network], thus covering a limited area" (Meeting environmental advisor, partner company) |

| Open knowing/data is politically charged | Mobilize stakeholder | "[W]e can see fish [in our web portal], so imagine that the local fishermen can go in there and look: 'Is there a point in going to the sea today?'" (Interview, environmental advisor, NorthOil) "The Arctic area is an incredibly complex picture with too many actors" (Internal meeting, principal researcher, NorthOil) |
|---|---|---|
| | Ideology of neutral knowing/science in policy making | "The purpose of the provisions on impact assessments is to clarify the effects of plans and measures that can have significant impact on the environment and society. The handling of cases under the provisions shall help to take into account the effects on environment and society when planning an action and when it is considered whether the measure will be implemented. The rules will also ensure an open process so that all parties concerned are heard." (Statement by the Norwegian government[4]) "If [the ministry of climate and the environment] thinks that the definition of the ice edge is set scientifically, she needs to prepare to defend it politically." (Kagge and Brønmo 2015) |

---

[4]https://www.regjeringen.no/no/tema/plan-bygg-og-eiendom/plan--og-bygningsloven/plan/ku/konsekvensutredninger/id410042/

# 5 FINDINGS: SYNTHETIC KNOWING OF THE ARCTIC

## 5.1 Making the Marine Environment Accessible

The regulatory approach of the Norwegian government regarding oil and gas activities ostensibly underscores the knowledge-based underpinning of granting operating permissions:

> "Official decisions that affect biological, geological and landscape diversity shall, as far as is reasonable, be based on scientific knowledge of the population status of species, the range and ecological status of habitat types, and the impacts of environmental pressures."
> (Norwegian Ministry of the Environment 2009, Section 8 "Knowledge Base")

Proactively addressing environmental concerns and public perceptions, NorthOil has since the 2000s strengthened its capacity for marine environmental monitoring. This is particularly vivid with oil companies lobbying to expand into the Arctic.

Marine environmental monitoring has traditionally been an offline, time- and resource-consuming task. It is regulated by national or local authorities and conducted by environmental experts from third-party service companies every third year. Usually, the process consists of collecting samples, photographs, and measurements of water quality, sediments on the seabed and the benthic flora and fauna around an installation. Samples are taken onshore for laboratory analysis over 9-12 months.

NorthOil considered traditional monitoring practices as insufficient to satisfy environmental concerns about the threats posed by oil and gas activities. NorthOil thus worked towards an IoT-based marine environmental monitoring capacity for real-time feeding decision-making. NorthOil in 2011 embarked on the EnviroTime project to establish a "platform…to monitor and analyze the environment in parallel with daily operations in order to protect sensitive areas and to minimize the risk of potential negative impact on the environment" (internal documentation, NorthOil). EnviroTime was headed by NorthOil with an industrial consortium covering key areas of expertise: subsea technologies, IT integration, and environmental risk assessment.

The immediate challenge facing EnviroTime was what (and how) to capture the unbounded variety of the marine environment with IoT. We focus here on one thread, viz. capturing living matter in the water (plankton, fish, larvae and eggs) in the form of an algorithmic phenomena coined 'biomass' in EnviroTime. The EnviroTime platform was intended to monitor biomass using the IoT with real-time visualizations to relevant users, thus informing and supporting operational decision-making. An accompanying real-time algorithm would help assess environmental risk during operations.

The process of bounding the variety of the marine environment highlighted the necessarily scoped nature of NorthOil's efforts of measuring the environment remotely. In EnviroTime, such efforts were shaped by both practical and political constraints. With fishing being Norway's second largest export industry (after oil and gas), it was important for the IoT to 'see' commercially interesting fish such as cod, herring, mackerel, and salmon. One of the companies involved in EnviroTime, a market leader of subsea sensor technologies in Scandinavia, proposed to deploy state-of-the art acoustic sensors that send an acoustic wave and measure the echo returned when the wave hits a target, such as a fish (similar to radar). The acoustic devices in EnviroTime were thus placed on the sea floor, a few hundreds of meters below sea level. This choice illustrates the scoped nature of IoT-rendered marine environment. First, it is only the reflection of the air-filled swim bladder that allows fish to float, with which only some fish are endowed, that is perceptible hence visible to the acoustic sensor.

Second, data were scoped by the granularity and reach of the measurements offered by the available sensors. EnviroTime explored different frequencies and power levels of acoustic sensors. Larvae and eggs floating close to the sea surface hence outside the range of the acoustic sensors could not be captured. As a result, refraining from technical details, it suffices to say that fish without a swim bladder (e.g., mackerel) and eggs were initially invisible to EnviroTime.

The subsequent challenge for the EnviroTime project was to make the IoT-generated biomass measure 'real' in the sense that it shaped decision-making, a principal ambition of the project. For biomass to become more than a mere measurement, there had to be established a baseline in order to have insight into the range of normal variation. Against such a baseline, biomass measures above a certain threshold could for instance be used to halt operations if threatening fish spawning or migratory patterns. However, biomass measures vary dramatically spatially and temporally including seasonally. Fixing a baseline proved non-trivial, because valid historic datasets were difficult to come by.

**5.2 Fixing a Baseline: Historical and Open Datasets**

By 2013, all EnviroTime project partners recognized the dire need for historical datasets to calibrate the biomass measure. Incidentally, a few NorthOil environmental advisors involved in EnviroTime were also part of other marine environmental monitoring campaigns run by NorthOil. The Venus Ocean Observatory (Venus project for short) was particularly significant, with half a decade of environmental datasets. The Venus project, EnviroTime members hoped, could be used for the much-needed historical datasets.

The Venus project was a key part of NorthOil's lobbying for oil and gas activities, presently banned, in the Venus area. Since 2005, the Venus project explored a variety of IoT configurations to measure water quality, acoustic devices to detect the concentration of biomass, and subsea cameras to take pictures of a rare species of cold-water coral reef, of which the world's densest population is located on the Norwegian seafloor. The datasets were stored on hard disks for later retrieval.

NorthOil's resolve to push into the Arctic was spearheaded by their strategic, corporate Arctic Program. This program recognized in 2011 the political significance of the Venus project in promoting NorthOil's image of environmentally concerned. One NorthOil environmental advisor candidly explained the utility of the Venus project to us as follows:

> "[T]here is of course one reason why we are doing this: it is to gather background data for potential future [oil and gas] operations… [W]e don't know. But in the meantime, we have this observatory and we are going to use it for testing both software and hardware technologies."

With the recognition from the Arctic Program followed additional funding to turn the Venus project into a permanent marine IoT monitoring station approximately 20 km off the coast of northern Norway, with a fiber-optic cable connected to an onshore data center. The monitoring station consisted of a handful of sensors installed on a semi-conic metallic structure that measured pressure, temperature, salinity, cloudiness, and concentration of biomass in the water column, in addition to a camera (Figure 1). Completed in 2013, it provided the first openly available, IoT-based, real-time marine environmental datasets in Norway.

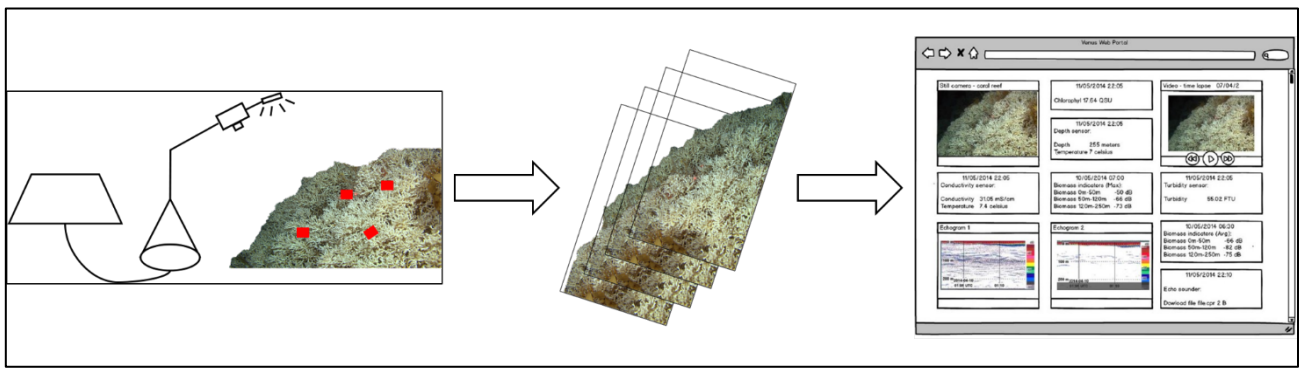

Figure 1. A schematic representation of the camera installed on a crane on the sea floor in Venus with the purpose of detecting marine resources in the proximity of a coral reef (red rectangles, figure on the left). Photographs are taken every 30 minutes, transferred via a fiber-optic cable (center), and are visualized via a web portal in real time (right). Reproduction by the authors. Photo credit: MAREANO/Institute of Marine Research, Norway.

The fishing industry has and still is a key antagonist to the oil and gas lobby pushing for the Arctic. NorthOil recognizes the political importance of forging alliances, when possible, with the fishing industry. The oil and gas lobby is eager to stress that collaborative relations with the fishing industry do exist. As one proponent underscores[5], "Fishermen have unique local knowledge and can mobilize at short notice. Their recruitment will provide a further strengthening of the oil spill preparedness organization near shore". Similarly, another representative of the oil and gas lobby stated, "We have already solved many challenges together with the fisheries. Among other things, we have done a lot together to find a time window for seismic shootings"[6].

The Arctic Program through the Venus project tried in multiple manners to enroll the interests of the fishermen, as we elaborate. The fiber-connected, IoT-based Venus Ocean Observatory needed a high-speed connection from the onshore data center, not only on the seabed. In collaboration with local fishermen, NorthOil thus decided to finance a fiber-optic Internet connection to the fishermen's village near the data center.

With the funding from the Arctic Program, the Venus project developed a web portal to store, manage, and visualize real-time measurements (Figure 2). The Venus portal was published under a Creative Commons license and openly accessible online, thus feeding an ideology of open, value-free scientific knowledge. The datasets are owned by NorthOil but may be downloaded for further use or publication, assuming due acknowledgment.

Equipped with the open Venus portal, NorthOil sought to make it relevant and useful to the fishermen, a key stakeholder in the politicized Arctic. As one NorthOil environmental advisor enthusiastically envisioned it, "[W]e can see fish, so imagine that the local fishermen can go in there and look: 'Is there a point in going to the sea today? Do [the fish] stay at home?'"

---

[5] Refer to statement by The Norwegian Oil and Gas Association, https://www.norskoljeoggass.no/en/News-archive/Miljo/Fishermen-to-join-oil-spill-preparedness-organisation/, accessed November 1, 2017.

[6] Fenstad, A., and Hagen, J. M. 2017. "Over halvparten av sjømatfolket sier olje-nei [More than half of people involved in the seafood industry are against oil]," Fiskeribladet, August 25. (https://fiskeribladet.no/nyheter/?artikkel=54912). Accessed November 1, 2017.

A workshop was organized in November 2013 to present the Venus portal to a community of fishermen. The feedback from the fishermen, rather unexpectedly, was quite positive according to our NorthOil informant present. A representative of the fishermen community commented that he wished that there were more such observatories in the area because they would be useful for predicting the amount of fish available to catch each day.

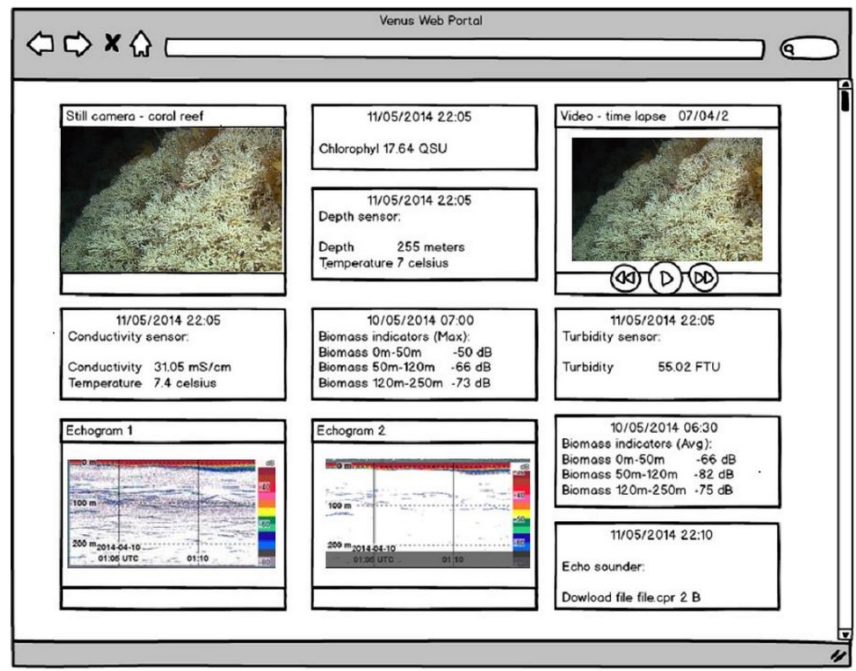

Figure 2. A mockup of the web portal developed as part of the Venus project. Photo credit: MAREANO/Institute of Marine Research, Norway.

**5.3 Tying in with the Politics of the Ice Edge**

The results of the Venus project attracted attention and visibility by 2013. The project demonstrated the technical feasibility of IoT based marine environmental monitoring. The Arctic Program recognized its potential political significance including the fact that "The Arctic area is an incredibly complex picture with too many actors" (Internal meeting, principal researcher, NorthOil). NorthOil was during this period strenuously lobbying for lifting the then-imposed ban on oil and gas activities in the Barents Sea, significantly further north than the Venus area[7]. A particularly challenging issue when pushing north, readily appreciated by NorthOil's Arctic Program, was the ice edge. Given the operational and environmental risks associated with ice[8], NorthOil's capacity for marine environmental monitoring had to be credible not only in the Venus area but also further north in the vicinity of the ice edge. In other words, the Venus project needed to generalize. As the head of the Arctic Program explained, using the proverbial, Norwegian 'potato' as a metaphor for generalizing, the ice edge is a proxy of generalizing as follows:

> "The Arctic Program wants to position [NorthOil] in the north, and the ice edge, in particular, is a very good potato: they need food to position [NorthOil] in the North (…) What is interesting for the Arctic Program is the issue of the ice edge, what resources are there, what can be visualized. There are many political aspects involved."

The Venus project collected, made accessible, and visualized a limited set of measurements from a confined spatial area. The challenge facing the EnviroTime project was accordingly to demonstrate NorthOil's more general capacity for marine environmental monitoring to make the push further north politically credible. In the EnviroTime project, NorthOil with partners addressed the adoption of the open data for predictive modelling along two avenues: enriching the biomass construct and working biomass into institutionalized methods for fish assessments in NorthOil. We elaborate.

---

[7] The lobbying succeeded. In the 23rd concessional round announced in 2016, 10 new licenses were opened for oil and gas activities, 3 of these in the Norwegian part of the Barents Sea. A coalition of environmental NGOs sued the Government claiming that the 23rd round went against the rights for a healthy environment written into the Constitution. The case lost in a verdict by the first level of courts in January 2018. The activity in the Venus area is still banned.

[8] Refer to U.S. Department of the Interior. 2013. "Review of Shell's 2012 Alaska Offshore Oil and Gas Exploration Program," Washington, DC. (http://www.doi.gov/news/pressreleases/upload/Shell-report-3-8-13-Final.pdf). Accessed March 1, 2015.

First, the biomass measure in the Venus project was bounded by the financial limitations in place before gaining the attention of the Arctic Program. Notably, the sensors deployed in the Venus project were off-the-shelf, low-end devices. For example, the acoustic sensors were, as pointed out above, unable to detect fish larvae and eggs floating close to the surface outside the response range of the sensors:

> "We can't, for instance, [measure] larvae and eggs in the upper water masses because this [sensor network] is on the bottom [at -250 m] and has a reach of about 50 meters."
> (Environmental advisor, EnviroTime partner company)

Basing EnviroTime on the Venus project data accordingly limited the type of IoT measurements down from EnviroTime's initially envisioned much broader spectrum of data types. The lack of data about larvae and eggs threatened EnviroTime's ambition and political significance. It implied a lack of data about future generations of fish affected by a possible oil spill. As a workaround, the EnviroTime project decided to use historical data from surface sampling stations provided by external research institutions to feed a predictive model based on algorithms from one of the partners to generate missing data about particle dispersion. In this manner, the historical datasets needed by EnviroTime to establish baseline conditions were enriched.

Second, using biomass as an indicator for marine environmental monitoring assumed having established baseline conditions. Here, the EnviroTime project used the Venus project data (actual and model-generated). To make biomass organizationally real in the context of oil and gas, it had to tie in with NorthOil's institutional practices and vocabulary of risk assessment familiar to professional groups, e.g., drillers and production engineers involved in daily operations. The challenge was to translate the results of the biomass-monitoring algorithm engineered in EnviroTime into the risk assessment language familiar to drillers and production engineers. Existing practices relied on a semaphore-like visualization of risk. EnviroTime decided to mimic this and worked on methods to visualize biomass concentration in the water.

Adapting an approach previously devised by the Norwegian Directorate of the Environment[9], the EnviroTime team formulated the environmental value, a number expressed in decibels that summarizes the hourly concentrations of biomass in large cubes of the water column (e.g., from -50 m to -100 m). The environmental value is obtained by collapsing multiple of the original sections scanned by the acoustic sensors into a single section (Figure 3).

Mimicking the semaphore-like visualization of risk from control rooms, the EnviroTime participants developed a categorization structure based on five colors to classify the amount of biomass in the entire water column, summing up all the cubes in the water column. For example, the highest concentration was coded in red and indicated the highest risk probability for the marine environment, indicating a halt in operations.

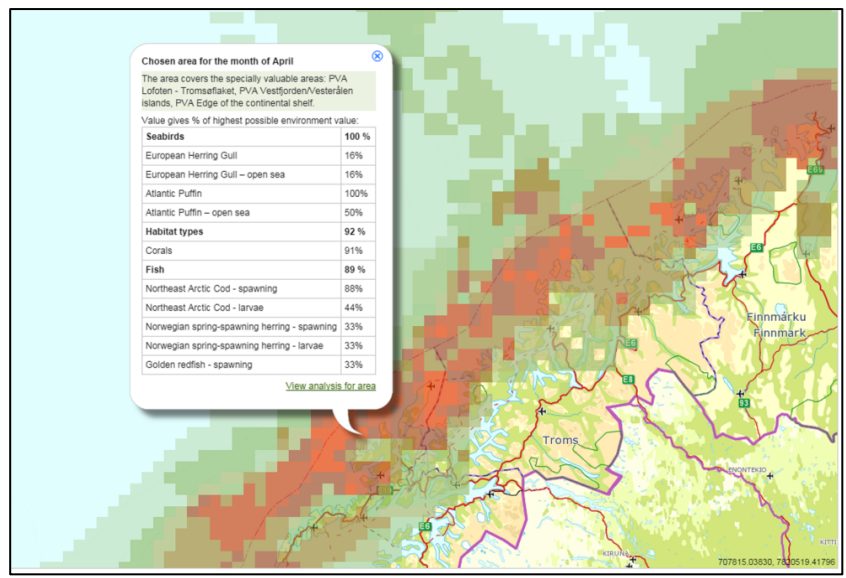

Figure 3. A portion of the Norwegian continental shelf where all offshore areas are mapped based on their corresponding environmental values (source: Norwegian Directorate of the Environment, havmiljø.no).

---

[9] Environmental values in Norwegian marine areas promoted by the Norwegian Directorate of the Environment: www.havmiljø.no. Accessed March 2, 2015.

# 6 DISCUSSION: TOWARDS SYNTHETIC KNOWING IN IoT-SATURATED CONTEXTS

All knowing is material (Orlikowski 2006). The challenge in IS research is to specify *how* knowing is material by drawing on theoretical characterizations of the digital (cf. Nambisan et al. 2017, p. 224). As Orlikowski and Iacono (2001) noted, there is a dire need to go beyond generic theorizing imported from other disciplines and start engaging with the specifics of the digital (cf. Monteiro and Hanseth 1996). Synthetic knowing, we propose, is knowing informed by theorizing digital materiality. We characterize and discuss synthetic knowing in the IoT-saturated context of the politically charged area of the Arctic.

First, *the objects of synthetic knowing become algorithmic phenomena*. Liquefaction is literally the capacity of digital representations to be unhinged from their physical referent (i.e., physical object, property or process); digital representations 'stand in for' physical referents (Bailey et al. 2012). This capacity underpins the potential for disruptive breaks with existing practices (Barrett et al. 2015; Lusch and Nambisan 2015; Nambisan et al. 2017). Unhinged, digital representations are amendable to subsequent algorithmic manipulations that fuel the recombination underpinning digital innovation (Henfridsson et al. 2018, Yoo et al. 2010). Increasingly self-referential, they gradually become the phenomena (Kallinikos 2007), or, more precisely, they become algorithmic phenomena (Orlikowski and Scott 2016). In our case, the physical phenomenon of the ice edge is gradually supplemented if not replaced by the algorithmic phenomenon of the ice edge. The physical ice edge is step-by-step superimposed by digital representations of the ice edge: measurements, initially from observations but later from satellite imaging, stand in for the physical ice edge. Subsequent algorithmic manipulations of these measurements through model-based, predictive simulations, for all practical purposes related to policy decisions about where to draw the line for oil and gas exploration activities, become the ice edge.

As Latour (1999) keenly observed, the qualitative richness of the physical phenomena may be traded for increased mobility and versatility of digital representations. Digital representations are never the same as the physical phenomena but, for given purposes, may fill their role. Knowing the ice edge relative to claimed knowledge-based policy decisions about oil exploration rely on the measurement- and simulation-based digital representations of the ice edge; synthetic knowing is knowing of the algorithmic phenomenon of the ice edge, with the physical ice edge increasingly out of sight.

The capacity for liquefaction is certainly but a potential. Whether digital representations carry organizational weight and whether they are made sufficiently credible to base consequential actions and decisions on remain empirically open issues (cf. Bailey et al. 2012, Dodgson et al. 2007). In and of themselves, digital representations are empty (or, in the vocabulary of Peirce's semiotics employed by Bailey et al. 2012: symbols). Digital representations gain weight, if at all, through social processes. The example of biomass in our case is illustrative. Biomass is an algorithmic phenomenon inferred from measurements of (a sample of) fish together with algorithmic manipulations. Eggs and larvae are not visible to the sensors and hence are generated by predictive models. In an effort to make biomass practically 'real', NorthOil attempted to accommodate biomass to the knowing practices of local fishermen. Despite initial enthusiasm, as reported in our findings, there are few if any fishermen today actively using biomass in their fishing practices, thus illustrating the organizational challenge of making algorithmic phenomena real in practice.

Second, *real-time streams increasingly conjure up phenomenological reality*. Sensors add a crucial dimension to the above noted general capacity of digital representations for liquefaction. Sensors *are* liquefaction vehicles. The range of physical objects, qualities, and processes measurable to sensors is wide and rapidly expanding (Singh et al. 2014). To illustrate from our case, the list of physical objects, qualities and processes targeted for sensor-based monitoring consisted of water currents, pressure, temperature, conductivity, turbidity, $O_2$, $CO_2$, oil-in-water, methane, chlorophyll fluorescence, topography, and benthic communities. In short, sensors capture an expanding richness of physical objects, qualities and processes. Sensors thus move closer to approximating human perception, hearing, tasting, smelling and tactile sensation, the backdrop of embodied knowing (Orlikowski 2006, Zuboff 1988).

In addition to capturing increased richness, sensors' ability to feed *real-time* streams make IoT-rendered digital representations compellingly 'real' (cf. notion of 'nowcasting' in Constantinou and Kallinikos 2015). As Knorr Cetina (2009, p. 72) notes, the real-time tickers of traders in her case create a 'fluid' reality, a synthetic situation where the digital representations (ticker feeds) are perceived as real "as the [real-time] information scrolls down the screens and is replaced by new information, a new market situation—a new reality—continually projects itself".

Taken together, the richness and the real-time capacity of sensor feeds increasingly conjure up the phenomenological lifeworld. The immediacy makes IoT-rendered representations phenomenologically real in manners in which for instance simulations are not. As underscored above, this capacity is never unbounded but always relative to a given purpose. The capacity of conjuring up phenomenological reality in real-time defines visions about virtual reality (VR) thus providing an opportunity to test the limits of this capacity. For instance, Prentice (2013, p. 83) studies how the embodied, tactile knowledge work of surgeons is replaced by digitally rendered substitutes, where "[real-time] graphics [of surgery] replace the sense of 'hands-on'", thus demonstrating Knorr Cetina's (2009) ontological fluidity of synthetic knowing practices. Certainly not in general, but for certain purposes such as education, the richness and immediacy of digitally rendered 'reality' is sufficient for the meaningful training of surgeons.

Third, *synthetic knowing is scoped*. With synthetic knowing, what you know is intractably connected with how you know it. This resonates with the so-called performative aspect of knowing (Cecez-Kecmanovic et al. 2014; Orlikowski and Scott 2008; Østerlie et al. 2012). Also Knorr Cetina's (2009) notion of scope captures this. A scope, Knorr Cetina (ibid., p. 64) explains, akin to a periscope is "an instrument for seeing or observing", which, in the case of traders, amount to "collecting, augmenting, and transmitting the reality of the markets" relative to the configuration of the scoping mechanisms. Within the domain of environmental monitoring, the relevance of scoping in synthetic knowing is visible through the configuration and selection of instruments and risk methods (e.g., worst-case vs most-likely scenarios) (Hauge et al. 2014; Knol 2013).

Our case demonstrates the scoped nature of synthetic knowing. The value of the biomass is an assigned decibel value for every cube of water in the water column. Biomass is a scoped construct. It depends on the commercial interests of NorthOil in addition to technical constraints. Regulators in Norway have traditionally focused more on fish than on sea mammals. Accordingly, biomass is constructed to attract attention and feedback from fishermen. The fish species monitored are the most commercially relevant ones.

Regardless of choice of echo sounder frequency, the fish that the sensors 'see' are the ones with a swim bladder. As it is really the bladder, not the fish, that provides the acoustic reflections for the echo sounders to read off, fish without a swim bladder or small eggs and larvae are literally not perceptible with echo sounder-based, scoped synthetic knowing of the biomass of NorthOil. The measured parameters of the biomass hence are supplemented with a predictive model (algorithm) for estimating the amount of eggs and larvae.

Fourth, *open knowing/data is politically charged*. Traditionally, marine environmental monitoring produced largely black-boxed results inasmuch for all practical purposes neither the datasets nor the details of the methods of analysis were disclosed. NorthOil's initiative for open marine environment data runs counter to this. Relying on the generative, open-endedness capacity of the digital that we have identified above, the push for 'open' environmental data by NorthOil feeds into and gains legitimacy from broader initiatives towards open data and data-driven knowing (Ruppert 2015).

Open environmental datasets underpin and legitimize policy decisions and "political agenda setting, allowing for the participation and empowerment of particular actor groups, possibly at the expense of others" (Lamers et al. 2016). There are several international efforts to promote open, marine environmental data motivated by the fact that "[w]e know more about the moon than the ocean floor"[10]. For instance, Ocean Observatories Initiative in the U.S. is a program that combines scientific platforms and distributed sensor networks that measure physical, chemical, geological, and biological parameters from the seafloor to the air-sea interface[11]. The Mareano program developed by the Norwegian Institute for Marine Research publishes detailed maps of the topography, geology, sediment composition, biodiversity, and pollution on the Norwegian continental shelf[12].

The political connotation of the push for open data/knowing stems in no small manner from its forceful rhetorical appeal to data as neutral or value-free. The strong and explicit ideological commitment to 'knowledge-based' political decision processes in Norway (Norwegian Ministry of Environment 2011), as Barret et al. (2013) point out, black-boxes the processes whereby data is produced, processes that demonstrate the anything but neutral character of data. In our case, the controversy around the ice edge illustrates the limitations of approaching open data as neutral data. With datasets and modeling in principle open, there is a fierce, ongoing debate around how to set the baseline for the ice edge and to frame natural variation. Environmental activists and NGOs together with a minority in Parliament advocate a long time-series going back to historic records of the ice edge (when there was more ice). The government, however, favors the selection of more recent data[13]. The government has challenged the existing definition of the ice edge, arguing that the available time-series measurements are outdated[14]. This demonstrates, in the colorful language of Edwards (1999, p. 452), the 'incestuous' relationship between a simulation model and the available data: one part of the data are used to build the model, the other parts to validate it.

---

[10] See Haugan, I. 2015. "Vi kjenner månen bedre enn havbunnen [We know the moon better than the seabed]," *Gemini.no*, May 5. (http://gemini.no/2015/05/vi-kjenner-manen-bedre-enn-havbunnen-var/). Accessed May 29, 2015.

[11] See http://oceanobservatories.org/. Accessed May 4th, 2016.

[12] Norway-based Mareano program, see http://www.mareano.no/en/. Accessed May 4th, 2016.

In the politicized context of Arctic oil, the openness of NorthOil's Venus web portal was used to mobilize diverse stakeholders such as politicians, local municipalities, fishermen, and the oil and gas lobby. The rhetorical appeals to openness, however, highlight the potential but not necessary the realization of openly accessing, analyzing and critiquing the data and their data-driven consequences (Ruppert 2015). As Shaik and Vaast (2016), drawing on Barrett et al. (2013), demonstrate, there more often than not is a significant gap between espoused and in-practice openness. For example, NorthOil exercises control over the open dataset by regulating how external parties may develop applications[15].

# 7 CONCLUSIONS

Knowing is material; hence, what we know and how we know it are entangled (Orlikowski and Scott 2008; Østerlie et al. 2012). To push forward in IS research by detailing *how* knowing is material, we need to theorize digital materiality (Nambisan et al. 2017; Orlikowski and Iacono 2001). Synthetic knowing is material knowing informed by theorizing digital materiality. With IoT, the range of objects of knowing (the algorithmic phenomena) expand with the expanding scope of what sensors are able to capture as built into visions of Industry 4.0 and Second Machine Age.

---

[13] See Kagge and Brønmo (2015). For additional details, refer to Norwegian Polar Institute. 2014. Iskant of Iskantsone - Fremstilling Av Iskantsonen Som Sårbart Areal [Ice Edge and Ice Edge Zone - Presentation of the Ice Edge Zone as a Vulnerable Area], Norwegian Polar Institute. (http://www.npolar.no/npcms/export/sites/np/no/fakta/iskant-ressurser/nedlastbart/iskant-og-iskantsone.pdf). Accessed January 15, 2018.

[14] See Koranyi, B. 2015. "Norway Redefines Arctic Ice Edge in Potential Boost for Oil Exploration," Reuters. (http://www.reuters.com/article/2015/01/20/norway-oil-arctic-idUSL6N0UZ0L020150120). Accessed March 1, 2017.

[15] In the vocabulary of digital platforms and ecosystems (Tiwana 2013), NorthOil takes on the role of a platform owner while inviting external, third party 'apps'.

If, as we suggest, the capacity of IoT-saturated reality is expanding the scope of synthetic knowing, this begs the question of the boundary conditions (Suddaby 2010). Contrary to the outlook of the most enthusiastic proponents (Chen et al. 2012; McAfee and Brynjolfsson 2012), the emerging picture we advocate is not one in which quantity trumps quality. However, where to draw the limit for synthetic knowing is empirical rather than ideological. It should invite neither dismissal nor uncritical appraisal but rather critical, empirical study (cf. Agarwal and Dhar 2014; Lycett 2013). That said, our analysis is based on an IoT-saturated situation – digital representations have a dominant if not constitutive role – that differs from many other work contexts. Our theoretical characterization of synthetic knowing is drawn from on one domain and set in a Scandinavian context, thus implications take the form of transferability rather than traditional generalization based on representativeness. In this light, our analysis of synthetic knowing should be seen as outlining a tendency many contexts of knowing are moving towards but have yet to have arrived at. Yet with processes of digitalization transforming industries and businesses significantly if unevenly, for many it need not be a distant tendency.

## ACKNOWLEDGEMENTS

The paper has had a long journey. Concepts and perspectives part to its development have grown out of numerous seminars and informal discussions. Their inception and evolution into the paper is difficult to account for but valuable ideas were provided by: Petter Almklov, Bendik Bygstad, Karin Knorr Cetina, Samer Faraj, Ole Hanseth, Vidar Hepsø, Georg von Krogh, Neil Pollock, David Ribes, Emil Røyrvik, and Robin Williams.

Editors and reviewers at the MISQ, very much including the SE, have consistently provided constructive feedback throughout the process.

Our research has been supported by the Research Council of Norway (NFR): IO Center (grant no. 40122563), Digital Oil (grant no. 213115), and Sirius (grant no. 237898).